\documentclass[twocolumn,trackchanges]{aastex63}
\usepackage{longtable}
\usepackage{xstring}
\usepackage{colordvi}
\usepackage{pmboxdraw}

\accepted{\today}
\submitjournal{ApJ}

\shorttitle{QUBRICS z$\sim$4 Luminosity Function}
\shortauthors{Boutsia et al.}
\begin{document}

\title{The Luminosity Function of bright QSOs at $z\sim 4$ and implications for the cosmic ionizing background}

\correspondingauthor{Konstantina Boutsia}
\email{kboutsia@carnegiescience.edu}

\author[0000-0003-4432-5037]{Konstantina Boutsia}
\affil{Las Campanas Observatory, Carnegie Observatories, 
Colina El Pino, Casilla 601, La Serena, Chile\\}

\author[0000-0002-5688-0663]{Andrea Grazian}
\affil{INAF--Osservatorio Astronomico di Padova, 
Vicolo dell'Osservatorio 5, I-35122, Padova, Italy\\}

\author[0000-0003-4744-0188]{Fabio Fontanot}
\affil{INAF--Osservatorio Astronomico di Trieste, 
Via G.B. Tiepolo, 11, I-34143 Trieste, Italy \\} 
\affil{IFPU--Institute for Fundamental Physics of the Universe, via Beirut 2, I-34151 Trieste, Italy}

\author[0000-0003-0734-1273]{Emanuele Giallongo}
\affil{INAF--Osservatorio Astronomico di Roma, Via Frascati 33, I-00078, Monte Porzio Catone, Italy}

\author[0000-0002-4096-2680]{Nicola Menci}
\affil{INAF--Osservatorio Astronomico di Roma, Via Frascati 33, I-00078, Monte Porzio Catone, Italy}

\author[0000-0002-7738-5389]{Giorgio Calderone}
\affil{INAF--Osservatorio Astronomico di Trieste, 
Via G.B. Tiepolo, 11, I-34143 Trieste, Italy \\}

\author[0000-0002-2115-5234]{Stefano Cristiani}
\affil{INAF--Osservatorio Astronomico di Trieste, 
Via G.B. Tiepolo, 11, I-34143 Trieste, Italy \\}
\affiliation{INFN-National Institute for Nuclear Physics,  
via Valerio 2, I-34127 Trieste \\}
\affil{IFPU--Institute for Fundamental Physics of the Universe, via Beirut 2, I-34151 Trieste, Italy}

\author[0000-0003-3693-3091]{Valentina D'Odorico}
\affil{INAF--Osservatorio Astronomico di Trieste, 
Via G.B. Tiepolo, 11, I-34143 Trieste, Italy \\}
\affil{Scuola Normale Superiore, 
P.zza dei Cavalieri, I-56126 Pisa, Italy\\}
\affil{IFPU--Institute for Fundamental Physics of the Universe, via Beirut 2, I-34151 Trieste, Italy}

\author[0000-0002-6830-9093]{Guido Cupani}
\affil{INAF--Osservatorio Astronomico di Trieste, 
Via G.B. Tiepolo, 11, I-34143 Trieste, Italy \\}
\affil{IFPU--Institute for Fundamental Physics of the Universe, via Beirut 2, I-34151 Trieste, Italy}

\author[0000-0003-4740-9762]{Francesco Guarneri}
\affil{INAF--Osservatorio Astronomico di Trieste, 
Via G.B. Tiepolo, 11, I-34143 Trieste, Italy \\}
\affil{Dipartimento di Fisica, Sezione di Astronomia, Universit\'a di Trieste, via G.B. Tiepolo 11, I-34131, Trieste, Italy}

\author[0000-0000-0000-0000]{Alessandro Omizzolo}
\affil{Specola Vaticana, Vatican Observatory, 
00122 Vatican City State\\}
\affil{INAF--Osservatorio Astronomico di Padova, 
Vicolo dell'Osservatorio 5, I-35122, Padova, Italy\\}

\begin{abstract}
Based on results by recent surveys, the number of bright quasars at redshifts z$>$3 is being constantly revised upwards. Current consensus is that at bright magnitudes ($M_{1450}\le -27$) the number densities of such sources could have been underestimated by a factor of 30-40\%. 
In the framework of the QUBRICS survey, we identified 58 bright QSOs at 3.6$\le z
\le $4.2, with magnitudes $i_{psf}\le$18, in an area of 12400$deg^{2}$. The uniqueness of our survey is underlined by the fact that it allows us, for the first time, to extend the sampled absolute magnitude range up to $M_{1450}= -29.5$. We derived a bright-end slope of $\beta=-4.025$ and a space density at $<M_{1450}>=-28.75$ of 2.61$\times 10^{-10} Mpc^{-3}$ comoving, after taking into account the estimated incompleteness of our observations. Taking into account the results of fainter surveys, AGN brighter than $M_{1450}=-23$ could produce at least half of the ionizing emissivity at z$\sim$4. Considering a mean escape fraction of 0.7 for the QSO and AGN population, combined with a mean free path of 41.3 proper Mpc at $z=3.9$, we derive a photoionization rate of $Log(\Gamma [s^{-1}])=-12.17^{+0.13}_{-0.07}$, produced by AGN at M$_{1450}<-18$, i.e.
$\sim$100\% of the measured ionizing background at z$\sim$4.

\end{abstract}


\section{Introduction} \label{sec:intro}

Studying the quasar (QSO) and active galactic nuclei (AGN) populations
at high-z is overly important for a number of reasons. Primarily, a 
quantitative estimate of their space density at different luminosities 
can give constraints on theoretical models aiming to
predict the formation and evolution of super massive black holes (SMBHs) in
the distant past \citep{Volo2020}. A detailed census of AGN, 
both at bright and 
faint magnitudes at $z>3$, can give interesting constraints on the sources 
responsible for the cosmological reionization of neutral hydrogen
\citep{Giallo15,Giallo19} and singly ionized helium \citep{Wor19}. 
In addition, the study of absorbers along the line of sight to bright 
QSOs at high-z can give precise information on the physical properties of 
the inter galactic medium (IGM, see references in \cite{calderone19}).
Moreover, the SMBHs, ubiquitous at the center of galaxies with bulges,
could be responsible, during their active phase, of the strong negative 
feedback that is able to suppress the star formation, eventually quenching 
the galaxy itself \citep[e.g.][]{Fiore17} and enriching with metals the 
circum galactic medium \citep{Travascio20}. Last but not the least, 
QSOs and AGN in general give a major contribution to the cosmic 
X-ray background and an important, though probably not dominant, 
contribution to the infrared (IR) background \citep{shen20}.

One of the most studied and important observational indicators for
the evolution of the AGN population is the QSO luminosity function, i.e.
their space density as a function of luminosity and redshift $\Phi(L,z)$.
In the past, the hunt for high-z QSOs has been limited to bright
magnitudes and selected areas of the sky, mainly based on photographic
plates in the optical \citep{SchGr83,KooKr88} or X-ray \citep{Boyle93} 
and Radio \citep{Gregg1996}.
The advent of wide area CCD detectors on
dedicated 2-4 meter class telescopes, i.e. the Sloan telescopes,
at the turning of the millennium \citep{fan2000} has allowed the 
massive search for high-z QSOs at relatively bright ($i\le 21$) 
optical magnitudes, breaking the record barrier of $z=6$ with a 
large sample of new QSOs
\citep{fan06}. At the present time, IR detectors allowed to extend the search
for the most distant and luminous QSOs at $z>7$ (e.g.  \citealt{Banados18,WangF18,YangJ20}).

At bright magnitudes ($M_{1450}\le -26$), the SDSS survey constituted an
unprecedented milestone for the space density of bright QSOs at $z\ge 3$ 
for at least 20 years, thanks to the thousands of newly discovered QSOs
at high-z \citep{Lyke2020}.
Most of the first studies on the AGN populations at high-z have been based 
on SDSS. Results based on these first studies, remained unchallenged 
until the Extremely Luminous QSO Survey \citep[ELQS,][]{Sch19a,Sch19b,Schindler17}. 
In this survey a combination of optical and IR colors has been used to 
select QSOs candidates through a supervised 
machine learning algorithm. This resulted into high completeness in bright 
magnitudes and an increase by 36\% of the known QSO population, in the 
targeted redshift range ($2.8\le z\le 4.5$). This suggests that the SDSS space densities at $z>3.5$ and magnitudes brighter than $M_{1450}=-27$ could be underestimated by a factor of 30-40\%.

The ELQ survey covers all the northern hemisphere, and extends to slightly
negative declination, but it does not cover the entire southern sky,
where major observational facilities will be deployed in the future
e.g. the Extremely Large Telescope (ELT), the Square Kilometer 
Array (SKA), a site of the Cherenkov Telescope Array (CTA). 
A dedicated effort to fill 
this gap has been undertaken recently. A survey searching for the brightest 
QSOs in the southern hemisphere, dubbed QUBRICS (QUasars as BRIght beacons 
for Cosmology in the Southern hemisphere), produced a new sample of
hundreds of QSOs at very bright optical magnitudes at $z>2.5$ \citep{calderone19,Bou20}.

The study of the luminosity function of high-z AGN is a
highly debated topic. Current efforts are focused on constraining the QSO luminosity
function at $z>6$ \citep{Matsu19,Jiang2016,Yang2019}, and even 
close to $z\sim 8$ \citep{Mori20}, while the luminosity
function at $3<z<5$ is not settled yet, both at the bright and 
faint end \citep{shen20}. One of the major problems in the study of
high-z QSOs is the completeness level of the different surveys, which is
quite difficult to measure. Efficient selections of high-z QSOs (e.g. SDSS)
are not usually associated to high completeness level, as shown by
\citet{Sch19a,Sch19b}. The QUBRICS survey is an attempt to search
for the brightest QSOs with negative declination with a well defined selection
criterion, which is highly complete for relatively bright objects
\citep{Bou20}. In this paper we consider the first subset of
QSOs at $3.6<z<4.2$, from this survey, to study the luminosity function of QSOs at very bright UV 
magnitudes $M_{1450}\sim -29$, which require wide areas on the sky and are still not well studied at present. 
In addition, we focus on regions of the sky that have been unexplored by previous studies so far, including SDSS. 

In Section 2 and 3 we present the selection procedure of the sources and the method used for calculating the luminosity function (LF). 
In Section 4 we present the best fit parameters and discuss the contribution of the QSOs to the ionising background based on our results.
In Section 5 we compare the observed space density of $z\sim 4$ AGN with the predictions from theoretical models and finally in Section 6 there is a summary of our results.
Throughout the paper we adopt the $\Lambda$ cold dark matter ($\Lambda$-CDM) concordance 
cosmological model (H$_{0}$ = 70 km s$^{-1}$ Mpc$^{-1}$, $\Omega_{M}$ = 0.3, and 
$\Omega_{\Lambda}$ = 0.7). All magnitudes are in the AB system.

%%%%%%%%%%%%%%%%%%%%%%%%%%%%%%%%%%%%%%%%%%%%%%%%%%%%%%%%%%%%%%%%%%%%%

\section{Data} \label{sec:selection}
\subsection{QSO Selection}

In 2018 our team started the QUBRICS survey \citep{calderone19}. With the goal of selecting high-z QSOs candidates, we used publicly available data from several databases: i) Skymapper \citep[DR1.1,][]{Wolf18}; ii) Gaia \citep[DR2,][]{Gaia18}; iii) 2MASS \citep{2MASS} and iv) the WISE survey \citep{Wri10}.
We limited our magnitudes in the range between $14\le m_i\le 18$ in order to only select bright sources and the declination to $d<0^{deg}$ so that our sources are in the Southern Hemisphere.
Then a Canonical Correlation Analysis \citep[CCA, ][]{CCA} has been applied in order to select robust high-redshift and bright QSO candidates. This was accomplished by training the algorithm using all previously known sources with secure identification from the literature.

The recipe obtained through CCA training has been applied to the rest of the sample in order to predict a classification. Stars have been identified based on parallax and proper motion information by Gaia and 
make up for $\sim$83\% of the sample. For sources classified as QSOs, a redshift estimate has been obtained 
using the CCA as a regression algorithm. This allowed us to reject lower redshift QSOs
($z<2.5$). The final list of high redshift bright QSOs included 1412 candidates. A pilot survey led to the initial discovery of 54 QSOs with $z\ge 2.5$. For more details 
about the selection method please refer to \cite{calderone19} (hereafter Paper I).

\subsection{Spectroscopic Follow-up}

Based on the encouraging results of the QUBRICS pilot campaign, we have
undertaken a more systematic spectroscopic follow-up for confirming the
nature of more candidates. We have been awarded time to several
facilities including the Low Dispersion Survey Spectrograph (LDSS-3) at the Clay Magellan telescope, the Inamori Magellan Areal Camera and Spectrograph (IMACS) at the Baade Magellan telescope, Wide Field CCD (WFCCD) at the duPont telescope and the ESO Faint Object Spectrograph and Camera (EFOSC2) at the New Technology Telescope (NTT). We observed 511 sources,
managing to obtain secure classification and redshift determination for
432. Most of our confirmed sources ($\sim$52\%) were bright QSOs at
z$\geq$2.5, of which 15 were at a redshift z$>$4. Our main contaminants
were lower redshift QSOs or AGN ($z<2.5$, $\sim 38\%$), while the rest
were galaxies and stars ($\sim$10\%). 

Taking into consideration the results of this campaign, we have updated the training sample and rerun the classification algorithm, thus obtaining a more robust list of candidates. In fact, based on this self-learning approach, our completeness has improved and is currently $>90\%$, while the success rate is close to 70\%. More details about the results of the spectroscopic follow up can be found in \cite{Bou20} (hereafter PaperII). The current sample of QSO candidates, in the redshift range 3.6$<z<$4.2, is based on the most recent selection described in PaperII. 

In the period November 2020 - January 2021 and after the publication of PaperII, we continued the spectroscopic follow-up of our 3.6$<z_{cca}<$4.2 QSO candidates. Observations have been obtained with IMACS and LDSS-3 at the Magellan telescopes. On LDSS-3 we used the 1"-center slit with the VPH-all grism and no filter. This results to a wavelength coverage of 4000-10000 {\AA} at a R$\sim$900 resolution. In order to obtain a similar resolution with IMACS we used the \#300 grism at a blaze angle of 17.5$^{o}$ and the 1" slit covering a wavelength range of 4000-10000 {\AA}. The data reduction and calibration for both instruments have been done following the recipes presented in PaperII. The sources presented for the first time in this work are commented in Table \ref{tab:qsoz4}
as "new data". Currently, from our initial candidate sample with $3.6<z_{cca}<4.2$ only 15 sources remain without spectroscopic follow up, of which 2-3 are of high quality. Thus our sample can be considered spectroscopically complete. 

\section{Analysis, Methods} \label{sec:LFunction}

Table \ref{tab:qsoz4} contains the 58 QSOs of $3.6 \le z_{spec} \le 4.2$
and $i_{psf}\le 18.0$ in the QUBRICS footprint. 

The ELQS survey by \citet{Sch19a,Sch19b} includes other QSOs in the same
redshift and magnitude interval, falling in the QUBRICS area but that are
not listed in Table \ref{tab:qsoz4}. The main reason is that those QSOs
have magnitudes $i_{psf}>18.0$ in Skymapper DR1.1, while the i-band
magnitudes by \citet{Sch19a,Sch19b} have been drawn from SDSS and 
Pan-STARRS1 \citep[PS1, ][]{PS1} photometry.

The QSO 015041-250846 by \citet{Sch19b} at z=3.600 (id=7250804)
is not included in our sample since, based on our data, we calculated a spectroscopic
redshift of $z=3.596$ (Paper II). This is slightly lower than our
redshift cut for the luminosity function calculations.

Three QSOs from the literature: 58209836 and 58674889 from \citet{Sch19a} and BRI 1117-1329 from \citet{StoLom1996}, were not part 
of our sample, due to incomplete photometry in the Skymapper
and WISE databases. We checked a posteriori that they 
have $i_{psf}\le 18.0$ and fall on the QUBRICS footprint, so  in principle, we should include them in our sample. In practise, we decided to compute the luminosity function of $z\sim 4$ AGN
by using only objects from our main sample, with the appropriate completeness corrections. The result would have been approximately the same if we had included them in
our calculations, but had neglected the incompleteness correction. Thus, only sources, presented in Table \ref{tab:qsoz4}, 
were used to compute the luminosity function of QSOs at $3.6\le z_{spec}\le 4.2$ at the bright end, i.e. $M_{1450}\le -28.0$. 

\begin{table*}
\caption{The $3.6\le z_{spec}\le 4.2$ QSOs at $i_{psf}\le 18.0$ in the QUBRICS Survey.}
\label{tab:qsoz4}
\begin{center}
\begin{tabular}{c c c c c c c}
\hline
$ID_{Skymapper}$ & RA & Dec & z$_{spec}$ & $i_{psf}$ & M$_{1450}$ & Reference \\
DR1.1 & J2000 & J2000 & & AB & & \\
\hline
68291629 & 11:35:36.40 & +08:42:19.08 & 3.847 & 17.966 & -27.915 & DR14$^{a}$ \\          
65558414 & 12:49:57.26 & -01:59:28.76 & 3.665 & 17.960 & -27.809 & ELQS$^{b}$ \\ 
57913424 & 11:49:14.40 & -15:30:43.97 & 4.129 & 17.756 & -28.289 & PaperI \\ 
65911949 & 13:20:29.98 & -05:23:35.29 & 3.700 & 17.444 & -28.346 & %Veron10 
\citet{Mitchell1990}\\       
56483517 & 09:35:42.70 & -06:51:18.93 & 4.040 & 17.424 & -28.570 & PSELQS$^{c}$ \\ 
113197224 & 16:16:48.96 & -09:14:44.39 & 4.055 & 17.876 & -28.126 & PSELQS \\ 
114286192 & 16:21:16.93 & -00:42:50.87 & 3.703 & 17.386 & -28.405 & ELQS \\ 
10623942 & 03:05:17.92 & -20:56:28.12 & 3.960 & 17.960 & -27.988 & PSELQS \\ 
68092164 & 11:30:10.59 & +04:11:28.12 & 3.930 & 17.718 & -28.212 & %Veron10 
\citet{Schneider2005}\\       
56662952 & 09:40:24.13 & -03:23:04.07 & 3.900 & 17.630 & -28.282 & PSELQS \\ 
58206167 & 10:14:30.28 & -04:21:40.31 & 3.890 & 17.571 & -28.336 & PaperI \\ 
57936842 & 10:20:00.81 & -12:11:51.49 & 3.715 & 17.904 & -27.897 & PSELQS \\ 
98382043 & 15:23:12.41 & -16:27:22.92 & 4.120 & 17.977 & -28.063 & PSELQS \\ 
13303827 & 04:11:02.07 & -01:35:15.10 & 3.660 & 17.911 & -27.854 & PSELQS \\ 
135386798 & 20:03:24.11 & -32:51:45.05 & 3.783 & 17.296 & -28.545 & %PKS 2000-330 
\citet{Peterson1982}\\ 
7766951 & 01:03:05.51 & -24:49:25.20 & 3.865 & 17.758 & -28.135 & PSELQS \\ 
7437380 & 00:03:22.95 & -26:03:18.17 & 4.111 & 17.071 & -28.963 & %Veron10 
\citet{Sargent1989}\\        
5533851 & 23:09:59.27 & -12:26:02.91 & 3.730 & 17.863 & -27.946 & PSELQS \\ 
8489172 & 01:13:51.96 & -09:35:51.17 & 3.668 & 17.875 & -27.895 & DR14 \\           
9182350 & 02:16:46.94 & -09:21:07.21 & 3.675 & 17.762 & -28.013 & DR14 \\           
10165846 & 01:50:48.82 & +00:41:26.31 & 3.703 & 17.970 & -27.821 & %Veron10 
\citet{Trump2006}\\       
8566706 & 01:40:49.17 & -08:39:42.40 & 3.713 & 17.635 & -28.163 & ELQS \\ 
8937029 & 02:21:23.90 & -14:16:54.87 & 3.650 & 17.753 & -28.005 & PSELQS \\ 
8430815 & 01:03:18.06 & -13:05:09.89 & 4.072 & 17.242 & -28.770 & PaperII \\
136588662 & 20:11:58.77 & -26:23:40.86 & 3.657 & 17.662 & -28.102 & PaperII\\
58181076 & 10:51:22.70 & -06:50:47.82 & 3.810 & 17.345 & -28.513 & PaperI \\
57143774 & 10:52:21.62 & -19:52:37.95 & 3.660 & 17.741 & -28.024 & PaperI \\
57929040 & 10:15:29.37 & -12:13:14.23 & 4.190 & 17.255 & -28.824 & PaperI \\
10739949 & 04:07:45.29 & -32:15:37.84 & 3.750 & 17.693 & -28.128 & PaperI \\
10934139 & 04:50:11.37 & -43:24:29.75 & 3.946 & 17.798 & -28.142 & PaperI \\
57933437 & 10:15:44.12 & -11:09:22.80 & 3.865 & 17.485 & -28.408 & PaperI \\
302866544 & 19:18:57.68 & -65:44:52.38 & 3.842 & 17.848 & -28.029 & PaperI \\
135100950 & 19:53:02.67 & -38:15:48.40 & 3.712 & 17.305 & -28.492 & PaperI \\
136198132 & 20:17:41.49 & -28:16:29.83 & 3.685 & 17.388 & -28.394 & PaperI \\
2379862 & 21:25:40.96 & -17:19:51.32 & 3.897 & 16.548 & -29.363 & PaperI \\
57368436 & 10:54:49.69 & -17:11:07.36 & 3.750 & 17.107 & -28.714 & PaperI \\
5528935 & 23:08:27.03 & -13:32:56.21 & 3.830 & 17.736 & -28.134 & PSELQS \\ 
317253125 & 00:48:05.34 & -59:29:09.44 & 3.607 & 17.536 & -28.196 & PaperII \\
317343050 & 01:27:16.87 & -58:02:47.28 & 3.918 & 17.772 & -28.152 & PaperII\\
58723356 & 11:13:32.47 & -03:09:13.98 & 3.731 & 17.949 & -27.860 & PaperII \\
315607762 & 03:17:24.89 & -57:36:19.01 & 3.844 & 17.922 & -27.956 & PaperII \\
310206031 & 05:09:43.13 & -74:09:47.89 & 3.773 & 17.575 & -28.260 & PaperII \\
316591563 & 05:29:14.28 & -45:08:07.03 & 3.690 & 17.661 & -28.123 & PaperII \\
14930439 & 04:36:23.92 & -00:04:02.89 & 3.852 & 17.404 & -28.479 & PaperII \\
%\hline 
\end{tabular}
\end{center}
\end{table*}

\begin{table*}
\caption{Table 1 - continued}
\begin{center}
\begin{tabular}{c c c c c c c}
\hline
$ID_{Skymapper}$ & RA & Dec & z$_{spec}$ & $i_{psf}$ & M$_{1450}$ & Reference \\
DR1.1 & J2000 & J2000 & & AB & & \\
\hline
305336573 & 21:08:17.67 & -62:17:57.53 & 3.794 & 17.589 & -28.259 & PaperII \\
317411112 & 00:18:30.46 & -53:35:35.20 & 3.738 & 17.744 & -28.070 & PaperII \\
316292063 & 05:48:03.20 & -48:48:13.19 & 4.147 & 16.886 & -29.169 & PaperII \\
307536920 & 21:51:37.44 & -44:36:44.17 & 3.638 & 17.363 & -28.388 & PaperII \\
6932623 & 02:04:13.26 & -32:51:22.80 & 3.835 & 17.068 & -28.807 & PaperII \\
8789744 & 01:55:58.27 & -19:28:48.98 & 3.655 & 17.393 & -28.370 & PaperII \\
4045023 & 21:55:13.29 & -03:16:05.61 & 3.690 & 17.410 & -28.374 & PaperII \\
6986244   & 02:35:57.55 & -34:48:56.45 & 3.737 & 17.792 & -28.022 & new data \\
10444829  & 04:08:28.43 & -39:00:32.93 & 3.610 & 17.817 & -27.916 & new data \\
10331020  & 03:12:52.40 & -31:38:33.21 & 3.879 & 17.828 & -28.072 & new data \\
309271177 & 02:10:51.46 & -84:54:37.57 & 3.685 & 17.170 & -28.609 & new data \\
305864039 & 23:34:54.76 & -69:30:42.84 & 3.894 & 17.856 & -28.052 & new data \\
316874745 & 03:27:24.51 & -52:38:58.20 & 3.771 & 17.787 & -28.047 & new data \\
60628332  & 12:11:20.09 & -33:14:27.46 & 3.826 & 17.728 & -28.141 & new data \\
\hline
\end{tabular}
\tablecomments{
 %     \small
      \\
      $^{a}$ DR14 refers to \cite{Paris18} \\
      $^{b}$ ELQS refers to \cite{Sch19a} \\
      $^{c}$ PSELQS refers to \cite{Sch19b} \\
      QSOs with $M_{1450}>-28.0$ are not used for the luminosity function calculations.
      }
\end{center}
\end{table*}

%\newpage

\begin{figure*}[h]
\plotone{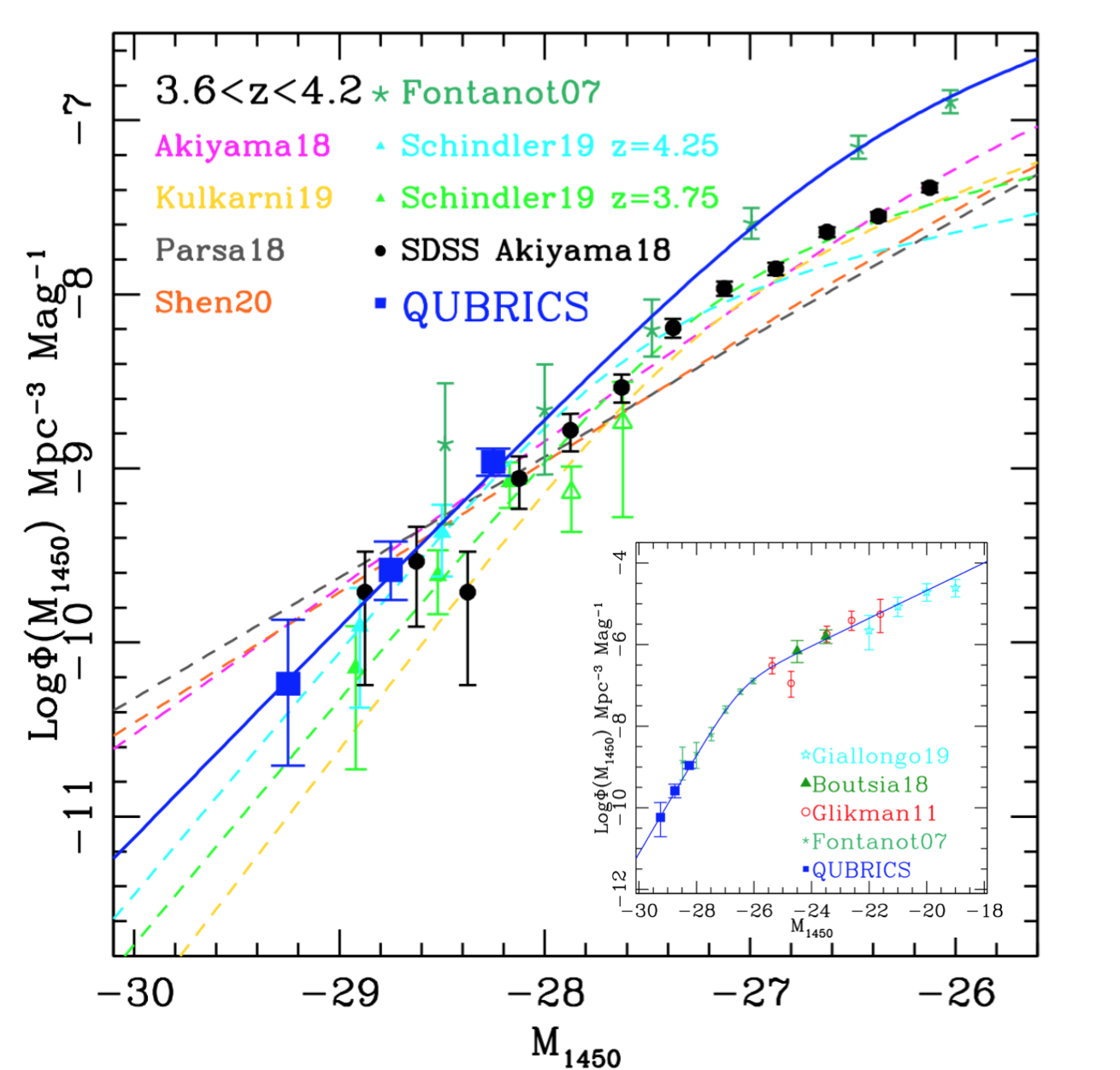}
\caption{The luminosity function of QSOs at $3.6\le z\le 4.2$ from QUBRICS (blue 
filled squares) compared to other luminosity functions from the recent literature.
All the data points and curves have been shifted to $z=3.9$ adopting the density evolution 
recipe by \citet{Sch19a} with $\gamma=-0.38$. The best fit result is shown by the blue line. In 
the bottom-right inserted plot we show the best fit LF extended to faint magnitudes, as discussed 
in Section 4.1}
\label{fig:lfz4}
\end{figure*}

Absolute magnitudes at 1450 {\AA} rest frame 
($M_{1450}$) in Table \ref{tab:qsoz4} have been derived
from the apparent magnitudes $i_{psf}$ of Skymapper and from 
the spectroscopic redshifts with the equation
\begin{equation}
M_{1450} = i_{psf}-5log(d_L)+5+2.5log(1+z_{spec})+K_{corr} \, ,
\end{equation}
where $d_L$ is the luminosity distance in parsec (pc) and the
k-correction $K_{corr}$ is given by the expression
\begin{equation}
K_{corr}=-2.5\alpha_\nu log_{10}(\lambda_{obs}/(1+z_{spec})/\lambda_{rest}) \, ,
\end{equation}
where $\alpha_\nu=-0.7$ is the typical spectral slope of QSOs, 
$\lambda_{rest}=1450$ {\AA}, and $\lambda_{obs}=7799$ {\AA} is the central
wavelength of the $i_{psf}$ filter.

\subsection{Completeness corrections}

Correcting the AGN space density for possible incompleteness effects
is important for the comparison of the QUBRICS luminosity function
with the results of other surveys.
The completeness of the QUBRICS sample at $3.6<z_{spec}<4.2$ has three factors:
\begin{itemize}
\item
c1: Sources that are not part of the Main Sample of 1014875 objects,
which is the starting catalog of QUBRICS, as described in PaperI and PaperII.
\item
c2: Sources that are part of the Main Sample, but have not been selected
by the CCA or $z_{CCA}$ criteria of PaperI or PaperII.
\item
c3: QSO candidates that are still missing spectroscopic identification.
\end{itemize}

In our analysis, we can assume that c2=1.0, since in Table
\ref{tab:qsoz4} we provide all the confirmed QSOs that are part of the
Main Sample, regardless they have been selected by the criteria of
PaperI, PaperII, or by other surveys. This choice has been achieved in
order to be less dependent on the assumptions usually carried out in
completeness simulations, e.g. the QSO spectral slopes, the equivalent
width distribution of the emission lines, the IGM transmission, the
photometric noise of the employed catalogs.

Regarding the correction factor c3, only 15 QSO candidates with
$3.6<z_{CCA}<4.2$ are still missing spectroscopic identification, but among
them, we expect to find no new QSO at $z\sim 4$: indeed, after visual
inspection of their spectral energy distributions, we have preliminary
indications that they are probably not high-z QSOs. Only two sources have a
spectral energy distribution consistent with $z\sim 4$ QSOs. Thus we can safely
assume here that c3=1.0, with small uncertainties with respect to the measured
Poissonian errors of our sample.

Estimating the correction factor c1 is not an easy task. At
this aim, we start our analysis from 881 known QSOs with
$3.6<z_{spec}<4.2$ used in PaperI as a training set for our CCA
selection. In this case, we do not introduce the new QSOs discovered
by QUBRICS in this analysis. We cross-correlate these 881 QSOs with
the public catalog of Gaia EDR3 \citep{gaiaedr3}, restricting the
analysis to the area covered by QUBRICS and limiting the $R_P$
magnitude of Gaia at $R_P^{gaia}\le 17.67$, which corresponds to a
Skymapper i-band magnitude $i\le 18.0$, which is the main criterion
for our Main Sample. We verify that the adopted photometric cut is
consistent with a selection in absolute magnitudes $M_{1450}$ brighter
that -28.0, which is the fainter limit of our luminosity function. We
end up with 32 known QSOs with $3.6<z_{spec}<4.2$. We then cross-correlate these 32 objects
with our Main Sample, finding 27 sources. The 5 missing objects are
bright ($i<18.0$) QSOs, but they have not been selected in our Main
Sample, due to their photometric flags in Skymapper or WISE
surveys. Considering these numbers (27/32=0.844), the correction factor c1 is 1.185 (1/0.844), and it does not
depend on the Skymapper i-band magnitudes of the selected QSOs.

\begin{figure}[]
%\plotone{MabVSzPapv2.png}
\includegraphics[width=265pt]{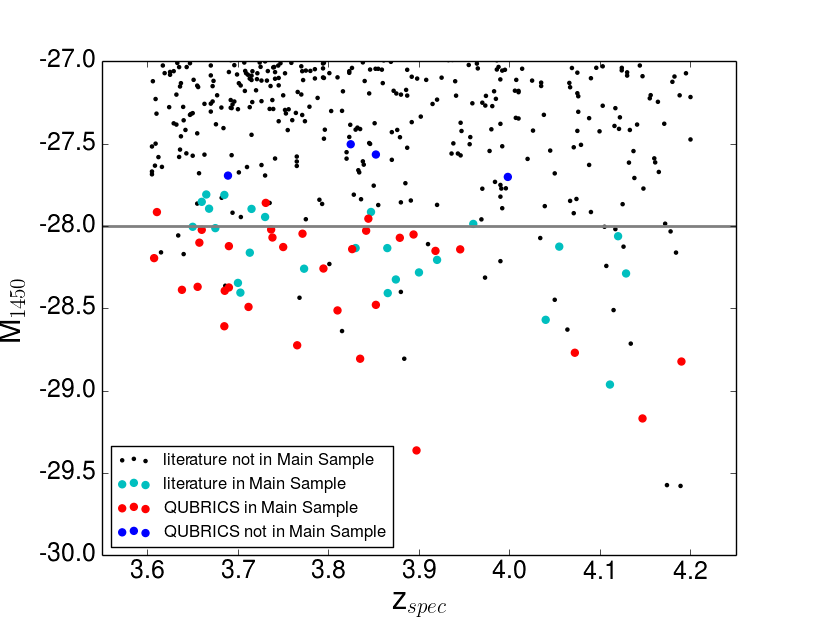}
\caption{Absolute magnitude M$_{1450}$ vs. spectroscopic redshift for all known QSOs in the redshift range 3.6$\le z_{spec}\le$4.2. Red symbols show the sources observed by QUBRICS and are part of the Main Sample. Blue symbols are sources observed by QUBRICS but are not part of the Main Sample. Cyan points show QSOs known from the literature that are also part of the Main Sample. Sources with M$_{1450} \le -28$ that are also part of the main sample (red and cyan points), have been used for calculating the luminosity function in this work. The line indicates the M$_{1450}$ = -28 magnitude limit.} 
\label{fig:Mab}
\end{figure}

Summarizing, we have applied a correction factor of 1.185 to the
space density of $z\sim 4$ QSOs of QUBRICS shown in Fig.\ref{fig:lfz4}. The absolute magnitude versus redshift of all known QSOs with  M$_{1450}<$-27 are shown in Fig.\ref{fig:Mab}. Only sources that are part of the main sample and have an absolute magnitude of M$_{1450}<-28$ (red and cyan points) have been used to calculate the luminosity function.

\section{Results: QSO Density Determination}

The space density of bright QSOs at $z\sim 4$ has been derived by adopting
the $1/V_{max}$ approach \citep{Eales1993}, where the accessible volume for each object has
been computed from the redshift interval encompassing $3.6\le z \le 4.2$ and the
magnitude limits of the survey. The accessible volume has been corrected by the factor
1.185 due to possible incompleteness of our survey, as discussed in the previous section.

The total area of the QUBRICS survey adopted here is 12400$deg^{2}$,
and the sample is limited to $14.0\le i_{psf}\le 18.0$ (Papers I and II). 
Only robustly confirmed QSOs (i.e. flagA) in the redshift
interval $3.6\le z\le 4.2$ have been used in the luminosity function calculation.
Error bars to the QSO space density have been computed by adopting the
statistics of \citet{Gehrels86}, which is particularly suited for low number
counts, while it is close to Poisson for large numbers.

We set the faintest absolute magnitude limit for the LF estimate to $M_{1450}=-28.0$, which is
the luminosity of an object at $i_{psf}=18.0$ at $z=4.2$, at the redshift limit of
our survey. This criterion includes 47 out of 58 sources presented in Table \ref{tab:qsoz4}.
We compute the QSO space density in three independent
intervals of absolute magnitudes, from the brightest limit $M_{1450}=-29.5$
to $M_{1450}=-28.0$. 
In Fig.\ref{fig:lfz4} we show the three independent bins.
Table \ref{tab:lfobs} summarizes the observed
space densities of bright QSOs found by QUBRICS in the redshift interval
$3.6\le z\le 4.2$.

A first consideration drawn from Fig.\ref{fig:lfz4} is that the bright
end of the luminosity function at $z\sim 4$ is rather steep, if compared to
previous results by SDSS \citep[e.g.][]{Fan2001,Rich2003}
and also by recent results \citep{Aki18,shen20}.

We have carried out a Maximum Likelihood analysis of the $z\sim 4$
QUBRICS QSOs following the formalism by \cite{Marshall83}. We
have fitted a single power-law with slope $\beta$, finding a best fit
value of $\beta=-4.14$ with a 68\% c.l. interval between -4.87 and
-3.56. This confirms the previous result of the ELQS survey by \cite{Sch19a} of
a relatively steep slope of the bright-end of the z=4 QSO luminosity function.

An attempt of fitting the LF with a double power-law has shown strong degeneracies between the bright end slope $\beta$ and and
$M^*$ (the absolute magnitude of the LF knee). This is expected since from previous works \citep{McGreer13,Sch19a}
it is known that this parameter is $M^*\sim -26$ at $z\sim 3-4$, much fainter than
our survey limit. Using
the QUBRICS sample presented in this paper, we are only able to put 1
$\sigma$ constraints of $\beta<-3.33$ and $M^*>-29.10$. This is
somehow expected given that our survey is limited to $M_{1450}=-28.0$,
much brighter than the expected break of the luminosity function of $z\sim 4$
QSOs. For these reasons, we decided to add fainter space densities
from the literature in order to provide a best fit analysis of all the
parameters ($\alpha$, $\beta$, $M^*$, $\Phi^*$) of the QSO luminosity
function, as we describe in the following section.

Another notable point of Fig.\ref{fig:lfz4} is the QSO space density at $M_{1450}=-29.25$,
which is a unique determination not available in other surveys.
This confirms the unique added value of the QUBRICS survey, and its success
in finding the most rare and brightest cosmic beacons, at least in the 
Southern hemisphere.

Our luminosity function determination is in agreement with the brightest points of
SDSS \citep{Aki18} and with \citet{Sch19a} at z=4.25. The error bars
of our data points are significantly smaller than the SDSS and the ones 
presented by \citet{Sch19a,Sch19b}.
Results from \citet{Sch19b} indicate that SDSS can be incomplete at
$\sim 40\%$ level, confirming previous values by \citet{Fontanot07}.
The results of the QUBRICS survey, shown in Fig.\ref{fig:lfz4}, seem to 
confirm such statements. 

\begin{table*}
\caption{The space density $\Phi$ of $3.6\le z\le 4.2$ QSOs in the QUBRICS footprint.}
\label{tab:lfobs}
\begin{center}
\begin{tabular}{c c c c c c}
\hline
\hline
Interval & $<M_{1450}>$ & $N_{QSO}$ & $\Phi$ & $\sigma_\Phi(up)$
& $\sigma_\Phi(low)$ \\
 & & & $cMpc^{-3}$ & $cMpc^{-3}$ & $cMpc^{-3}$ \\
\hline
$-28.5\le M_{1450}\le -28.0$ & -28.25 & 36 & 1.089E-09 & 2.136E-10 & 1.809E-10 \\
$-29.0\le M_{1450}\le -28.5$ & -28.75 &  9 & 2.611E-10 & 1.196E-10 & 8.581E-11 \\
$-29.5\le M_{1450}\le -29.0$ & -29.25 &  2 & 5.802E-11 & 7.712E-11 & 3.838E-11 \\
\hline
\hline
\end{tabular}
\end{center}
The space density $\Phi$ has been corrected for incompleteness, as discussed in the main text.
\end{table*}

\subsection{Best fit to LF data down to $M_{1450}=-18$}

In order to provide a best fit to the QUBRICS data on a wider magnitude 
range that covers both the bright and faint ends, we also considered 
luminosity function determinations at lower luminosities. 
For this analysis, we adopted a double power-law function for the LF as described below: 

\begin{equation}
    \phi = \frac{\phi*}{10^{0.4(M-M^{*}_{1450})(\alpha+1)}+10^{0.4(M-M^{*}_{1450})(\beta+1)}} \,.
\end{equation}

We include
in our best-fit analysis the data from \citet[][hereafter F07]{Fontanot07}, based on a
re-analysis of the SDSS survey at $z\sim 4$ with a revised selection function.
The refined completeness correction by F07 induces a steep space
density of QSOs
at $M_{1450}\sim -26$ which is higher than the one computed by \citet{Aki18} (but consistent to a 1$\sigma$ level up to $M_{1450}\sim -28$).

We decided not to use the \citet{Aki18} QSO LF in this work since we have
indications from other works that it could be underestimated. 
In \cite{Bou18}, e.g., it is shown that the AGN space density of
\citet{Aki18} is three time lower than the one in the COSMOS field at
$M_{1450}\sim -23$ and five times lower than the estimates by \cite{Glikman11} on the NDWFS and DLS fields, as also discussed in \cite{Giallo19}. 
\cite{Bou18} have also shown that this
discrepancy cannot be due to cosmic variance effects on the COSMOS,
NDWFS or DLS areas. Since the \cite{Fontanot07} luminosity function is
in better agreement with the results of \cite{Glikman11} and \cite{Bou18}, 
both of which have been based on
spectroscopically complete samples of $z\sim 4$ AGN, we decided to adopt for our purpose the F07 QLF, that covers the range of absolute magnitudes $-28<M_{1450}<-24$.

Going at fainter luminosities, we rely on the results by \citet{Glikman11}, \citet{Bou18} and \citet{Giallo19}.
The best fit of the luminosity function
has been carried out by a minimum $\chi^2$ analysis on the above
mentioned binned data points. The best fit result is shown in the bottom right inserted plot 
in Fig.\ref{fig:lfz4} and the best-fit parameters, together with their 
1$\sigma$ uncertainty ranges are summarized in Table \ref{tab:lffit}.

\begin{table}
\caption{The best fit parameters of the QSO luminosity function at
$3.6\le z\le 4.2$ in the QUBRICS footprint.}
\label{tab:lffit}
\begin{center}
\begin{tabular}{c c c c c c}
\hline
\hline
$\alpha$ & $\beta$ & $M^*_{1450}$ & $Log\Phi^*$ \\
\hline
$-1.850_{-0.250}^{+0.150}$ & $-4.025_{-0.425}^{+0.575}$ & $-26.50_{-0.60}^{+0.85}$
& $-6.85_{-0.45}^{+0.60}$ \\
\hline
\hline
\end{tabular}
\end{center}
The errors associated to the best fit parameters are at 68\% confidence level (1 $\sigma$).
\end{table}

At bright magnitudes, space densities are higher than previous fits
by \citet{Kulkarni19}, \citet{Sch19a} at z=3.75, and \citet{Aki18,shen20} at
$M\ge -28$. At fainter luminosities, also the best fit by \citet{Sch19a}
at z=4.25 is inconsistent with the observed data points of F07, and
all the previous results in the literature failed to reproduce the
observed data points, especially at the faint side.

Based on the updated fit provided in Fig.\ref{fig:lfz4}, we proceed
with the derivation of the ionizing background produced by bright QSOs,
and faint AGN, at $z\sim 4$.

\subsection{The ionizing background at $z\sim 4$ produced by QSOs and AGN}

\begin{figure}
\plotone{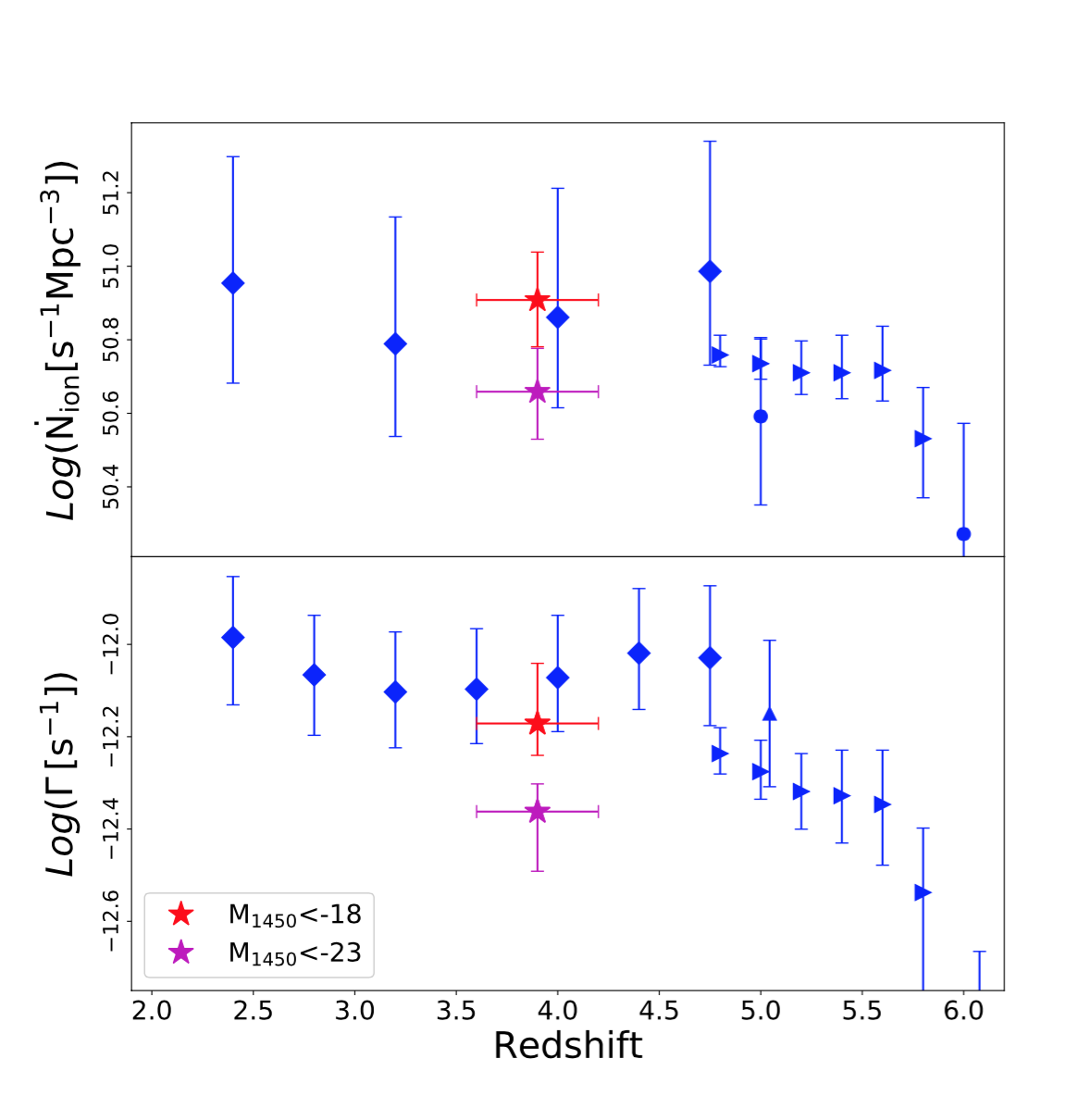}
\caption{Contribution of the AGN population to the ionizing background ({\it upper panel}) and to the photoionization rate ({\it lower panel}), based on our LF estimate, assuming a density evolution as in \citet{Sch19a} and a mean escape fraction (f$_{esc}$) of 0.7. The magenta star shows the contribution integrating to an absolute magnitude of -23 and the red star to -18. Error bars have been calculated as discussed in Section 4.2. Observed data from \citet[][circles]{WyitheBolton11}, \citet[][diamonds]{BeckerBolton13} and \citet[][triangles]{DAloisio18}.}
\label{fig:uvb}
\end{figure}

The detailed knowledge of the QSO luminosity function at z$\sim$4 can be used 
to estimate the AGN contribution to the photon volume emissivity 
($\dot {N}_{\rm ion}$) and 
photonization rate ($\Gamma$) (Fig.\ref{fig:uvb}). We apply the same formalism as in \citet{Fontanot14} and \citet{Cristiani16}: 

\begin{equation}
    \dot{N}_{ion}(z) = \int_{\nu_{H}}^{\nu_{up}} \frac{\rho_{\nu}}{h_{p}\nu}d\nu
\end{equation}

\begin{equation}
    \rho_{\nu} = \int_{L_{min}}^{\infty} f_{esc}(L,z) \Phi(L,z)L_{\nu}(L)dL \,,
\end{equation}

where $\rho_{\nu}$ is the monocromatic comoving luminosity density brighter that $L_{min}$, $\nu_{H}$ is the frequency corresponding to 912 {$\AA$} and $\nu_{up}$=4$\nu_{H}$.

The evolution of the photoionisation rate $\Gamma$ with redshift follows the parametrisation presented by \cite{HaardtMadau12}:

\begin{equation}
     \Gamma(z) = 4\pi \int_{\nu_{H}}^{\nu_{up}} \frac{J(\nu,z)}{h_{p}\nu}\sigma_{HI}(\nu)d\nu\,,
\end{equation}
where $\sigma_{HI}(\nu)$ is the absorbing cross-section for neutral hydrogen and $J(\nu,z)$ is the background intensity:
\begin{equation}
     J(\nu,z) = c/4\pi \int_{z}^{\infty} \epsilon_{\nu1}(z_{1})e^{-\tau_{e}} \frac{(1+z)^{3}}{(1+z_{1})^{3}} |\frac{dt}{dz_{1}}|dz_{1}\,,
\end{equation}

where $\nu_{1}$ is the proper volume emissivity and $\tau_{e}(\nu,z,z_{1})$ represents the effective opacity between z and $z_{1}$. 
Starting from the functional form we estimate for the QSO LF at $3.6<z<4.2$, we assume a pure density evolution consistent with the SDSS results \citep{Sch19a}, a bivariate distribution of absorbers as in \citet{BeckerBolton13}, 
an escape fraction $f_{\rm esc}=0.7$ \citep{Cristiani16} for all QSOs and a mean free path of 41.3 pMpc at z=3.9 \citep{Worseck14}. We then solve 
the equations of the radiative transport in a cosmological context \citep[see e.g][]{HaardtMadau12}, assuming two different luminosity limits for the QSO LF, corresponding only to the QSO contribution ($M_{\rm UV}<-23$) and to the 
total AGN population ($M_{\rm UV}<-18$). 

Following the considerations by \citet{DAloisio18}, we have increased the values of the ionizing 
emissivity by a factor of 1.2 to take into account the contribution by radiative recombination in the IGM. The resulting values are shown in Fig.\ref{fig:uvb} as red and magenta stars, and we collect the 
values in Table~\ref{tab:gamma}. 
The reported error bars (corresponding to the 16$^{th}$ and 84$^{th}$ percentiles of the distribution) are obtained by means of $\sim$25,000 Monte Carlo realizations varying: (a) the QSO LF parameters 
within 3 $\sigma$ confidence level defined by our minimization procedure, taking into account all relevant covariances; and (b) the intrinsic spectral slope of AGN emission between a single slope ($f_\nu \propto \nu^{\alpha}$ with $\alpha=-0.69$) and a broken power
law (with $\alpha=-1.41$ at $\lambda < 1000$ {\AA}). Our estimates are then compared with a set of observational determinations from \citet{WyitheBolton11}, \citet{BeckerBolton13} and \citet{DAloisio18}.

\begin{table}
\caption{The photon volume emissivity and photo-ionization rate per hydrogen atom produced
by bright QSOs and faint AGN at $z=3.9$.}
\label{tab:gamma}
\begin{center}
\begin{tabular}{c | c c c | c c c}
\hline
\hline
$M_{\rm UV}^{\rm lim}$ & \multicolumn{3}{c}{Log( $\dot{{\rm N}}_{\rm ion}$ $[s^{-1}$ $Mpc^{-3}]$)} &
 \multicolumn{3}{c}{Log( $\Gamma [s^{-1}]$)} \\
 & 50th & 16th & 84th & 50th & 16th & 84th \\
\hline
-23 & 50.66 & 50.53 & 50.78 & -12.36 & -12.49 & -12.30 \\
-18 & 50.91 & 50.78 & 51.04 & -12.17 & -12.24 & -12.04 \\
\hline
\hline
\end{tabular}
\end{center}
\end{table}

Based on Figures \ref{fig:lfz4} and \ref{fig:uvb} the following 
considerations can be drawn:
(i) the statistical errors on the luminosity function determinations
at z$\sim$4 are very small and at present no serious issue is present.
It is quite implausible that in the future new surveys of QSOs and AGN
at this redshift will change this picture dramatically, considering that our survey is spectroscopically complete.
(ii) Uncertainties on the photo-ionization rate $\Gamma$ are mainly
due to systematic effects, more precisely, the knowledge of the escape 
fraction of faint AGN and of the mean free path of QSO and AGN ionizing 
photons at $z\sim4$ \citep[e.g.][]{Romano19}.
In addition, measurements of the ionizing background are still uncertain 
by a factor of 2 (\citet{fg08} vs \citet{BeckerBolton13}). Improving such 
uncertainty would provide useful answers on the temperature of IGM.

Recent predictions presented by \citet{Dayal20} conclude that at 
z=4 AGN can provide a maximum of 25\% to the cumulative ionizing emissivity, considering a variety of models and escape fraction values. 
Their contribution could go as high as 50-83\% at z=5. 
Our observations of the photo-ionization rate indicate that AGN at z=4 and M$_{1450}<-23$ can provide more than 50\% of the UV background, 
and an even larger fraction if a different estimate for the ionizing background, like \citet{fg08}, is considered.
Reaching a sound conclusion on the role of AGN in the production of
ionizing photons in the post-reionization era needs a clarification on
the exact value of the ionizing background and on the mean free path
of HI ionizing photons at $z\sim 4$. At a lesser extent, the
measurement of the LyC escape fraction ($f_{esc}$) of faint AGN is another
important unknown in the present calculations.

If $f_{esc}$ of faint AGN is significantly below 70\%, then the calculations 
above are not
far from the real numbers: following \citet{Giallo19}, AGN fainter than
-23.0 are contributing only 10-20\% to the total ionizing background produced
by accreting SMBHs. The escape fraction of AGN at $M_{1450}\le -23$
turns out to be $\ge 70\%$ \citep{Cristiani16,Grazian18}, without any trend
with the observed optical luminosities. If the escape fraction is
rapidly dropping to zero at fainter magnitudes, then the total photo-ionizing
background would be lower by only 20\%, which is relatively small compared to
the bigger uncertainties still present on other quantities (the mean free path
and the UV background at present have uncertainties of the order of 50\%).
In the future it will be very important to derive with great accuracy
the value of the ionizing background at $z\sim 4$ \citep{fg08,BeckerBolton13} and the
mean free path of HI ionizing photons \citep{Prochaska09,Worseck14,Romano19}.
The LyC escape fraction of AGN fainter than $M_{1450}=-23$ is also important,
but not as fundamental, in the derivation of an accurate measurement for the
photo-ionizing background in the post reionization epoch.

%%%%%%%%%%%%%%%%%%%%%%%%%%%%%%%%%%%%%%%%%%%%%%%%%

\section{Discussion}

\subsection{Comparison with theoretical models}

\begin{figure}[]
\plotone{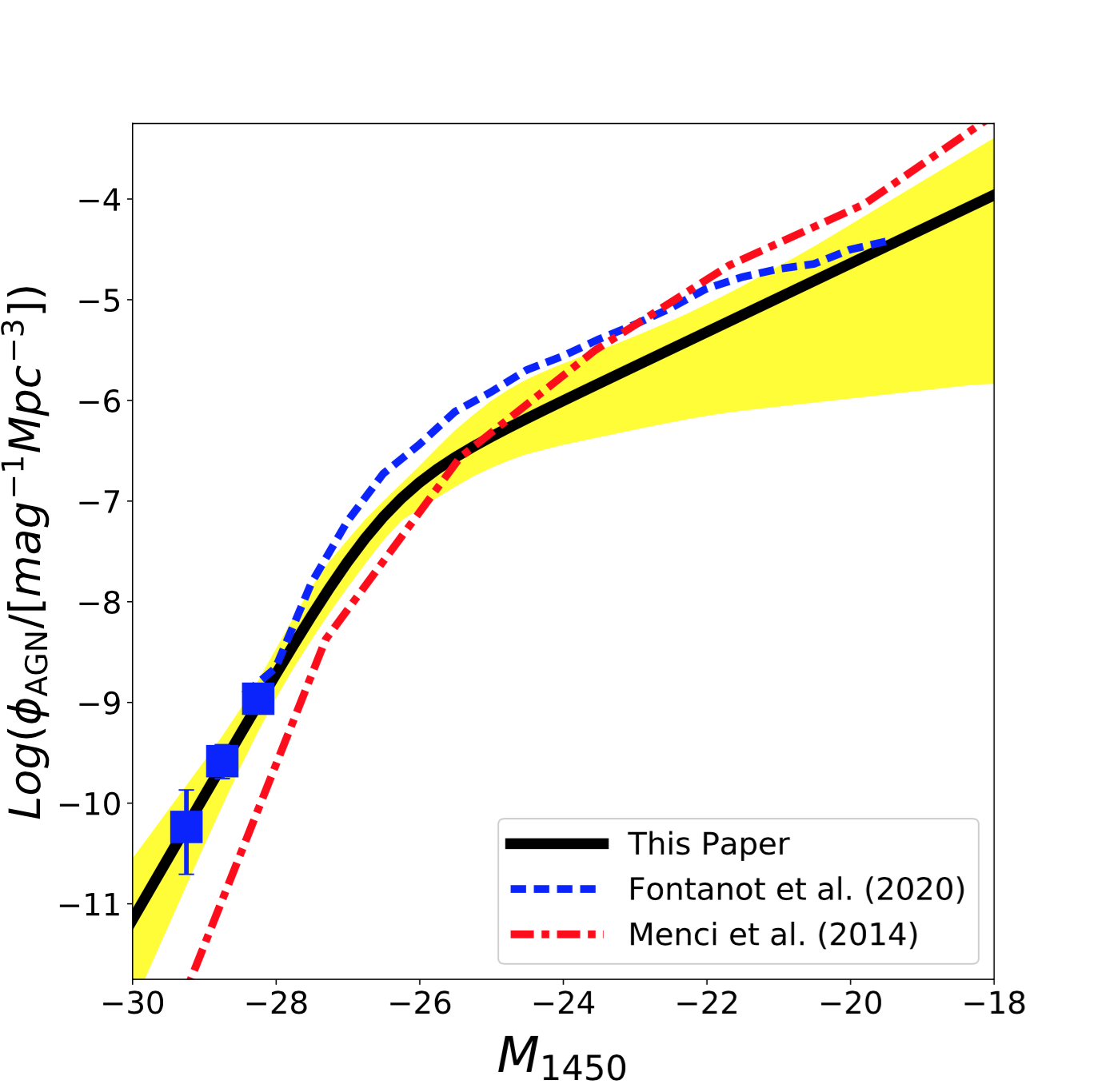}
\caption{Comparison of our best fit high-z LF with predictions from semi-analytic models. Shaded area represents the 3-$\sigma$ uncertainty range corresponding to the best fit parameters for the LF. Blue dashed and red dot-dashed lines correspond to predictions from {\sc gaea} \citep{Fontanot20} and the Rome SAM \citep{Menci14}, respectively.}
\label{fig:theoLF}
\end{figure}

We can compare the estimated high-z QSO LF with the predictions of theoretical models of galaxy formation and evolution. In particular, we consider the predictions of two semi-analytic models (SAM), the GAlaxy Evolution and Assembly model ({\sc gaea}, \citealt{Fontanot20}) and the Rome SAM \citep{Menci14}. These models are able to describe the formation and evolution of galaxies, starting from a statistical description of the Large Scale Structure and the distribution of the Dark Matter haloes, and assuming prescriptions (either empirically or theoretically motivated) to describe the key physical processes acting on the baryonic component. In particular, the models we consider in Fig.\ref{fig:theoLF} have been calibrated to reproduce the evolution of the bolometric QSO LF as described by optical \citep[e.g.][]{Hopkins07} and X-ray \citep[e.g.][]{Ueda14} surveys. 

Both models reproduce reasonably well our estimate of the QSO LF over the whole magnitude range (black solid line, shaded region represents the 3-$\sigma$ uncertainty range), with a possible overestimate of sources at $M_{1450}>-23$. The {\sc gaea} predictions are taken from a run based on merger trees extracted from the Millennium Simulation \citep[MS,][]{Springel05}. The MS volume is barely enough to sample 
the magnitudes corresponding to the fainter of our QUBRICS data (blue squares in Fig.\ref{fig:theoLF}), whose space density is in good agreement with model predictions. On the the other hand, the \citet{Menci14} SAM is based on merger trees following the Extended Press \& Schechter approach and is able to sample space densities corresponding to brighter magnitudes. The trigger for the BH accretion is provided by the minor and major interactions, in addition to the disc instability that should be of little influence in such bright QSOs. 
Nonetheless, at these luminosities this model predicts space densities slightly below the QUBRICS estimate. 

It is important to keep in mind that, despite these models have been explicitly calibrated to reproduce the evolution of the bolometric QSO LF, this effort typically focuses on the knee of the LF, where most of the sources lie. The bright-end of the LF, on the other hand, is populated by the most extreme objects, either in terms of accretion rate or SMBH mass, and Fig.\ref{fig:theoLF} clearly shows the relevant degree of uncertainty in this luminosity regime. These sources are indeed the most difficult to model, although they bring a lot of information on the evolution of structures in the early stages of structure formation.

\subsection{Additional considerations}

\begin{figure}[]
\plotone{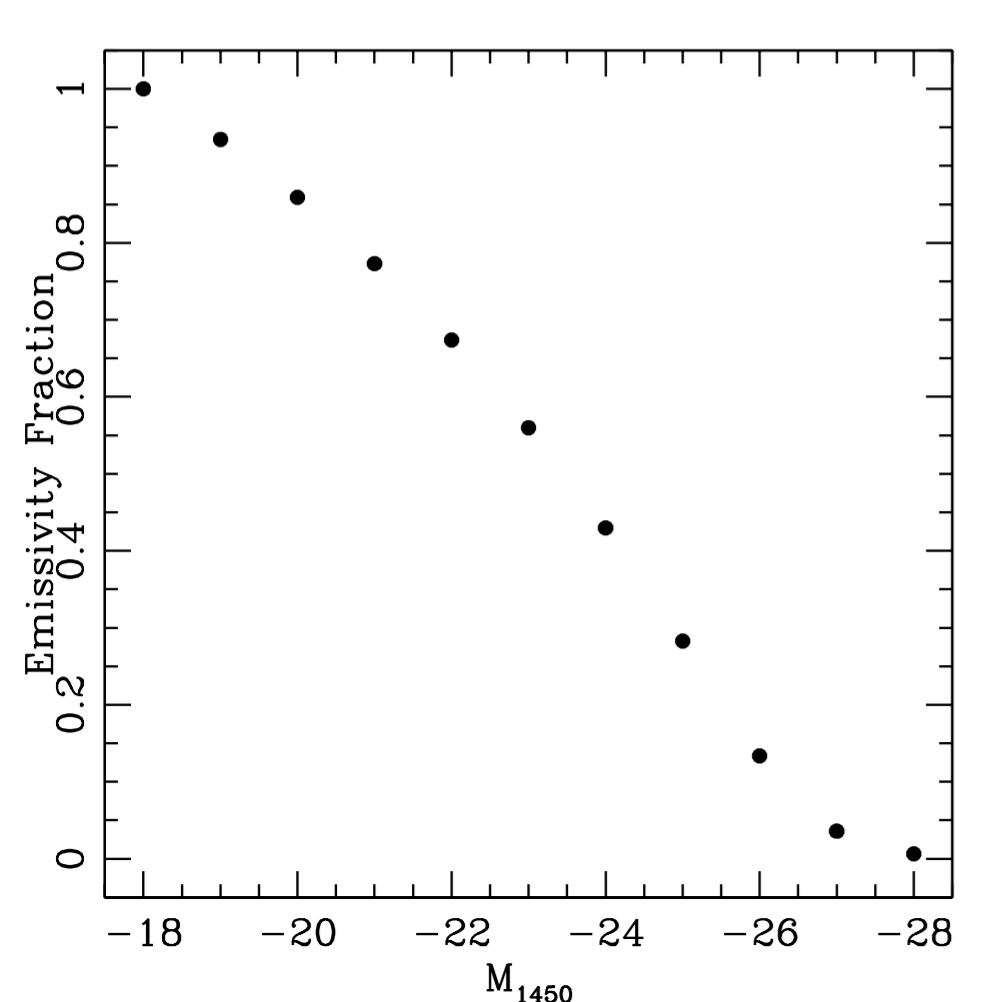}
\caption{Fraction of integrated ionizing emissivity at z$\sim$3.9.}
\label{fig:frac}
\end{figure}

Fig.\ref{fig:frac} provides the fraction of ionizing emissivity
$\epsilon_{912}$ produced by $z\sim 4$ QSOs at different luminosities.
The fraction is relative to the emissivity computed for M$_{1450} \leq$-18, according to the following equations:

\begin{equation}
%\begin{split}
 \epsilon_{ion}(z)=<f>\epsilon_{912}
%\end{split}
\end{equation}
and 
\begin{equation}
%\begin{split}
  \epsilon_{912}=\int_{}^{}\phi(L_{1450},z)L_{1450}\left(\frac{1200}{1450}\right)^{0.44}\left(\frac{912}{1200}\right)^{1.57}dL_{1450}\,
%\end{split}
\end{equation}

where $<f>$ is the average escape fraction from QSOs and $\epsilon_{912}$ is the ionizing emissivity produced by QSO activity.
The results are consistent with the one obtained by
\citet{Giallo19}, modulo the fact that our luminosity function is
slightly steeper, both in the bright and faint end, and
that the break of our fit is located at a slightly brighter
luminosity. The main reason for these differences is that in our fit
we did not consider the SDSS data by \citet{Aki18}, which are
lower than the ones by \citet{Fontanot07}, \cite{Glikman11} and \cite{Bou18}.

From Fig.\ref{fig:frac} it is easy to conclude that QSOs brighter
than $M_{1450}=-23.0$ provide at least half of the ionizing emissivity
at $z\sim 4$. If an escape fraction of 75\% is assumed for QSOs brighter
than $M_{1450}=-23.0$, and a lower escape fraction of 25\% is assumed
for the fainter population, then the total emissivity will result in 50.8\% 
($0.515\cdot 0.75+0.485\cdot 0.25=0.5075$). It is worth noting
that for $z\sim 4$ QSOs brighter than $M_{1450}=-23.0$, the measurements of
the escape fraction give values larger than 70\%
\citep{Cristiani16,Grazian18}, without any trend with the luminosity.
Thus, the scenario where sources fainter than $M_{1450}=-23.0$ will have
negligible escape fraction of ionizing photons is quite unrealistic. 
Interestingly, \citet{guaita16}
provided an example of an X-ray selected AGN of $M_{1450}\sim -21.9$
at $z\sim 3.5$ with an escape fraction of 72\%, and other AGN with lower
values. If confirmed by more extended samples, it can indicate that
also fainter AGN can have non negligible escape fraction of ionizing photons.

We can conclude from this analysis that the uncertainties on the
escape fraction of the faint AGN population can give a maximum uncertainty
of a factor of 2, with a photo-ionization rate which can be 2 times lower than
our value provided in Table \ref{tab:gamma}. This value has been derived
by adopting a mean free path of 41.3 pMpc by \citet{Worseck14}. If we instead
adopt a mean free path from \citet{Romano19}, which is $\sim$1.3 times larger,
then the resulting photo-ionization rate would be even higher. In practice,
the uncertainties on the escape fraction and mean free path can
compensate each other. In summary, the value of the photo-ionization rate
provided in Table \ref{tab:gamma} is robust with respect to the uncertainites
on the physical properties of the faint AGN population and the IGM.

At present, the measurement of the photo-ionizing background from the
literature, based mainly on the Lyman forest fitting, is uncertain by a factor
of 2 \citep{fg08,BeckerBolton13}. Considering that our estimate of $\Gamma$ is
also uncertain by a factor of 2, it can be concluded that our result does not support a scenario where AGN give a minor contribution to the ionizing UV background already at z=4 as proposed by e.g. \citet{Kulkarni19,Kim20}. 
Thus, it would be safe to conclude that, modulo the present 
uncertainites on the
escape fraction of faint AGN and the mean free path at $z=4$,
the QSO/AGN population alone can provide the amount of radiation to keep the
cosmological hydrogen fully ionized at these redshifts.

\section{Conclusions}

From the QSO luminosity function analysis at $z\sim 4$ and -29.5$<M_{1450}<$-28.0 from the QUBRICS survey it is possible to draw the following conclusions:
\begin{itemize}
\item
  Our $z\sim 4$ luminosity function extends to an unprecedented absolute
  magnitude of $M_{1450}=-29.5$. This result confirms the uniqueness of the
  QUBRICS survey in finding the most luminous objects in the distant Universe.
\item
  The best fit for the bright end slope of the $z\sim 4$ luminosity function is
  $\beta=-4.025$, significantly steeper than the slopes by
  \citet{Fan2001,Aki18,parsa18,shen20}.
  The expected slope
  is in agreement with the ones found by \citet{Sch19a,Sch19b}
  at similar redshifts. This implies that there is little evolution of the slope
  of the bright end of the QSO luminosity function from $z=1$ to $z=4$.
\item
  We find a higher space density of bright QSOs at $z\sim 4$ with respect to
  SDSS by a factor of 30-40\%, confirming previous results by
  \citet{Sch19a,Sch19b}. 
\item
  Our observed best fit QSO luminosity function at $z\sim 4$ is in tension with the results previously obtained by \cite{Fan2001,Aki18,parsa18,Kulkarni19,shen20}. More specifically, at bright magnitudes, the space density calculated by our survey is higher than all previously presented fits. Also at fainter magnitudes, most previous surveys failed to reproduce the observed data points by F07 and \cite{Bou18}, leading to underestimating the faint-end slope.
\item
  The HI ionizing background produced by bright QSOs and faint AGN (up to
  $M_{1450}=-18$) is $Log(\Gamma [s^{-1}])=-12.17^{+0.13}_{-0.07}$ (at 3 $\sigma$ confidence level), which is close to the value measured by
  \citet{BeckerBolton13} at similar redshifts. Our value has been derived by
  assuming a LyC escape fraction of 0.7 for the whole QSO and AGN population \citep{Grazian18}
  and a mean free path of 41.3 pMpc \citep{Worseck14}. If the mean free path calculated by 
  \citet{Romano19} is instead adopted, the photo-ionization rate produced by AGN
  can be higher by a factor of $\sim$1.3.
\end{itemize}

 The comparison of our observed QSO luminosity function at $z\sim 4$ with 
 the predictions of two semi-analytical models show that the extremely 
 bright-end probed by QUBRICS could give critical insights into the recipes 
 for triggering QSO activity in massive dark matter halos.
 In the future, it is important to extend the QUBRICS survey to deeper
regions of the sky in order to robustly determine the
location of the break, and the faint-end slope of the QSO luminosity
function at $z\sim 4$ and beyond, with a survey as
complete as possible. This attempt will require a large investment of
telescope time, but it will be feasible in the near future thanks to the large
imaging databases of the Vera Rubin telescope (LSST), the large field of view 
of the Roman Space Telescope (former WFIRST), and the powerful
spectroscopic capabilities of wide field spectrographs planned in the
next decades.

%---------------------------------------------------
\acknowledgments

AG and FF acknowledge support from PRIN MIUR project ‘Black Hole winds and the Baryon Life Cycle of Galaxies: the stone-guest at the galaxy evolution supper’, contract 2017-PH3WAT.

This work has made use of data from the European Space Agency (ESA) mission
{\it Gaia} (\url{https://www.cosmos.esa.int/gaia}), processed by the {\it Gaia}
Data Processing and Analysis Consortium (DPAC,
\url{https://www.cosmos.esa.int/web/gaia/dpac/consortium}). Funding for the DPAC
has been provided by national institutions, in particular the institutions
participating in the {\it Gaia} Multilateral Agreement.

This paper includes data gathered with the 6.5 meter Magellan Telescopes located at Las Campanas Observatory (LCO), Chile.

The national facility capability for SkyMapper has been funded through ARC LIEF grant LE130100104 from the Australian Research Council, awarded to the University of Sydney, the Australian National University, Swinburne University of Technology, the University of Queensland, the University of Western Australia, the University of Melbourne, Curtin University of Technology, Monash University and the Australian Astronomical Observatory. SkyMapper is owned and operated by The Australian National University's Research School of Astronomy and Astrophysics. The survey data were processed and provided by the SkyMapper Team at ANU. The SkyMapper node of the All-Sky Virtual Observatory (ASVO) is hosted at the National Computational Infrastructure (NCI). Development and support the SkyMapper node of the ASVO has been funded in part by Astronomy Australia Limited (AAL) and the Australian Government through the Commonwealth's Education Investment Fund (EIF) and National Collaborative Research Infrastructure Strategy (NCRIS), particularly the National eResearch Collaboration Tools and Resources (NeCTAR) and the Australian National Data Service Projects (ANDS)

This publication makes use of data products from the Wide-field Infrared Survey Explorer, which is a joint project of the University of California, Los Angeles, and the Jet Propulsion Laboratory/California Institute of Technology, funded by the National Aeronautics and Space Administration.

\facilities{Skymapper, Wise, Gaia, Magellan:Baade (IMACS), Magellan:Clay (LDSS-3)}

%------------------------------------------------------------------------------

\bibliography{ReferencesLF}{}
%% \bibliographystyle{aasjournal}

%\bibitem[Gehrels (1986)]{gehrels86} Gehrels, N. 1986, ApJ, 303, 336


\end{document}